Dynamic Infrared Electro-optic Response of Soluble Organic Semiconductors in Thin Film Transistors

Emily G. Bittle, Joseph W. Brill, Joseph P. Straley
Department of Physics and Astronomy, University of Kentucky, Lexington, KY 40506-0055 USA

**ABSTRACT**

We use a frequency-dependent electro-optic technique to measure the hole mobility in small molecule organic semiconductors, such as 6,13 bis(triisopropylsilylethynyl)-pentacene. Measurements are made on semiconductor films in bottom gate, bottom contact field-effect transistors (FETs.) Because of the buried metal layer effect the maximum response, due to absorption in the charge layer, will be for a dielectric film ~ 1/4 of a wavelength (in the dielectric) (e.g. ~ 1 micron thick in the infrared.) Results are presented for FETs prepared with both spin-cast polymer and alumina dielectrics prepared by atomic layer deposition. At low frequencies the results are fit to solutions to a non-linear differential equation describing the spatial dependence of flowing charge in the FET channel, which allows us to study multiple crystals forming across one set of drain-source contacts. FETs prepared on alumina dielectrics show interesting deviations from the model at high frequencies, possibly due to increased contact impedance.

**INTRODUCTION**

Reflecting the growing interest in developing devices using organic semiconductors and in understanding their properties, probes are being developed that can measure the spatial and dynamic dependence of semiconductor properties within the channel of thin-film field effect transistors, in which the mobile charge is induced by applied voltages. Optical probes have been used to measure the distribution and dynamics of charge in the channel [1-6] and are going beyond the usual "lumped impedance" approximation of the field-effect transistor (FET.)
In this study, we use spatially resolved voltage modulated infrared reflection to measure the dynamic response of charge in field effect transistors. We investigate the difference in transport between two dielectrics and the variation of mobility within one transistor. Results are compared to a model of dynamic response.

**EXPERIMENTAL DETAILS**

Samples are made with 6,13 bis(triisopropylsilylethynyl)-pentacene (TIPS-pentacene), provided by John Anthony's group [7], which is drop cast onto a transistor structure built on a glass slide. (Thermally oxidized silicon films have huge parasitic electro-optic responses, presumably due to absorbed species, so are inappropriate substrates for our measurements.) Drop casting the TIPS-pentacene results in large, thin crystals which can be oriented by tilting the sample during drying so that the long axis is parallel to the drain-source gap (L ≈ 350 microns.) Multiple crystals can form on one set of contacts.

The gate, drain and source electrodes are vapor deposited gold; the dielectric layer is either spin cast poly(4-vinyphenol) (PVP) cross-linked with polymelamine co-formaldehyde or alumina films, prepared by atomic layer deposition to avoid pinholes which typically occur in sputtered films. To maximize the absorption of infrared light (ν = 900-1000 cm$^{-1}$) in the conducting layers adjacent to the dielectric, the thicknesses of the dielectric layer were ~1 micron, which is ~ ¼ of the wavelength in the dielectric. [8]

Spatially resolved measurements, using an infrared microscope for measurement areas 30 x 30 μm, were taken of the change in reflectance while applying a gate voltage ($V_{gs}$) to a thin film transistor while holding the drain and source at ground. Infrared light, from a tunable lead-salt laser, is reflected off the gate (passing through the TIPS-pentacene twice), and absorbed by accumulated charges in the conducting layer resulting in a change in reflection ~10$^{-4}$. To gain information on the charging dynamics of the TIPS-pentacene transistor, a unipolar square wave voltage is applied to the gate at various frequencies, and we measured the relative change in reflectance, ΔR/R, both in-phase and in quadrature with the square wave, as functions of position (x) and amplitude, $V_{gs0}$. We assume that ΔR/R(x) α ρ(x), the induced charge. Additional details of the experiment can be found in Reference [1]. The frequency dependence of ΔR/R in the center of the channel was fit to solutions of

$$\frac{\partial \rho}{\partial t} = \left(D + \frac{\mu \rho}{C}\right)\frac{\partial^2 \rho}{\partial x^2} + \frac{\mu}{C}\left(\frac{\partial \rho}{\partial x}\right)^2 \quad (1)$$

where D is the Einstein diffusion constant, μ is the mobility, and C is the capacitance of the dielectric. Numeric solutions at various positions and frequencies are presented in Reference [8]. This one-dimensional equation assumes that fringing fields, leakage current, and contact impedances are negligible, so that ρ(0,t) = ρ(L,t) = CV$_{gs}$(t). We also assume D ~ μk$_B$T/e << μV$_{gs0}$. We can then solve for the mobility, which we assume is constant (μ$_0$) as discussed below.

**DISCUSSION**

**Dynamics and mobility variation on a single transistor**

Transistors grown from drop casting will often form long (~ 1 mm), wide (~ 0.1 mm) crystals, so that multiple "single" crystals can span one pair of drain and source contacts. These crystals often have different transport properties that are compounded in current-voltage measurements. Due to the spatial resolution of our system, we are able to measure the mobility of individual crystals. Figure 1 shows the change in reflectance signal as a function of the gate voltage frequency for three crystals on the same transistor and Table 1 gives the mobility for each crystal calculated from fits to the solution of Equation 1 as well as the linear mobility obtained from current-voltage (IV) measurements.

As seen in Figure 1 and Table 1, Crystal C is longer (which formed at ~45° off the drain-source gap) and therefore much slower than Crystals A and B (which formed nearly parallel to the drain-source gap.) The decreased mobility on Crystal C is possibly due to variation in the dielectric material, or due to the anisotropy in TIPS-pentacene conduction which would result in a factor of ~2 decrease in mobility for a crystal oriented at 45° [9]. The mobility measured over all transistors (IV) is most similar to the mobility of Crystal C, though the width of Crystal C is comparable to A and B and is therefore unlikely to dominate charge conduction. We consider the frequency dependent change in reflectance to be a more reliable measurement of mobility, since it relies on fewer parameters than mobility calculated using IV characteristics, and we therefore attribute the discrepancy to uncertainty (e.g. capacitance, bias stress effects) in the IV measurement.

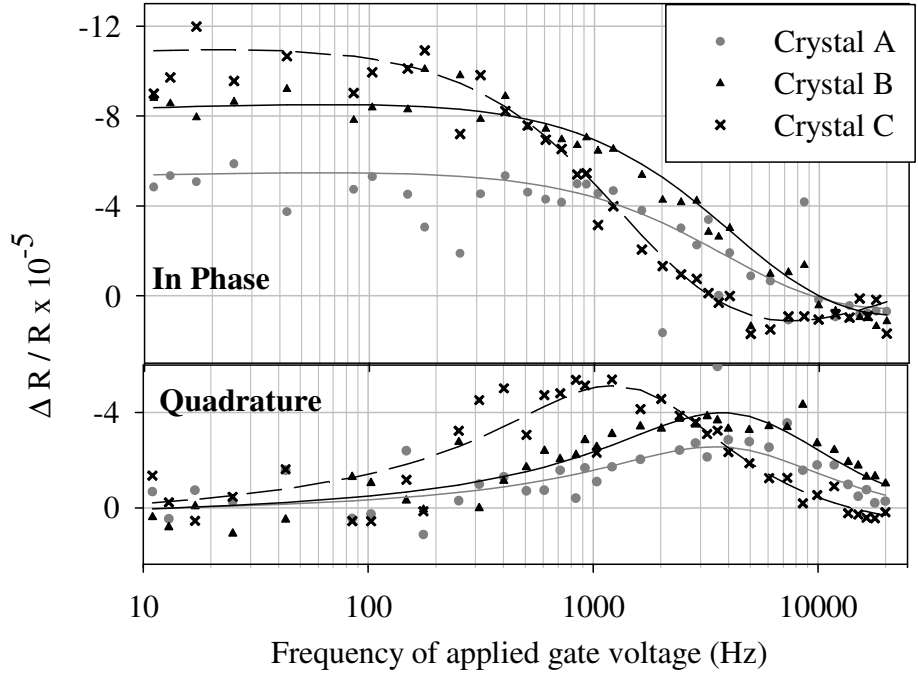

**Figure 1** Gate voltage frequency dependence of the change in reflectance at the center of a TIPS-pentacene transistor measured on three crystals spanning one drain-source contact pair, $V_{gs} = 0$ to -50 V. Fits (within error bars, not shown to avoid crowding) to the solution of Equation 1 are shown. Noise at lower frequencies (< 400 Hz) is larger than at high frequencies due to ambient electrical and mechanical noise.

**Table I** Mobilities calculated from the fits to data in Figure 1 and from current-voltage (IV) measurements. Uncertainty in mobility for crystals A, B, and C are ~20% of the final value, due to uncertainty in $V_{gs}$ (2%), L (5%), and $F_0$ (10%). $F_0$ and A are the fitting parameters for Equation 1, where the characteristic frequency $F_0 = (\mu_0 V_{gs})/(2L^2)$ and A is the amplitude of $\Delta R/R$, that depends on the charge density and interference effects. Length is measured from drain to source, and width is measured across the crystal.

| Crystal | $F_0$ (Hz) | A ($10^{-4}$) | Length (μm) | Width (μm) | Mobility (cm$^2$/Vs) |
|---|---|---|---|---|---|
| A | 2050 | 0.9 | 400 | 75 | 0.13 |
| B | 2200 | 1.4 | 375 | 125 | 0.12 |
| C | 720 | 1.8 | 500 | 100 | 0.07 |
| IV | - | - | 425 average | 300 | 0.07 |

### High frequency discrepancy on aluminum oxide

On samples prepared on alumina, the change in reflectance with frequency deviates from the calculated charge density at high frequency (Figure 2.) Dielectrics are known to affect the transport properties of organic FETs, due to increased polarization on high k dielectrics or increased trapping sites, among other effects. [10], [11] Charge trapping will cause a voltage dependence of the mobility. In comparing calculations of mobility dependent and independent charge density (inset, Figure 2), we find that this cannot explain the effects we are seeing. More likely we are seeing the effects of increased impedance at the contacts, possibly due to defects near the electrodes. [12]

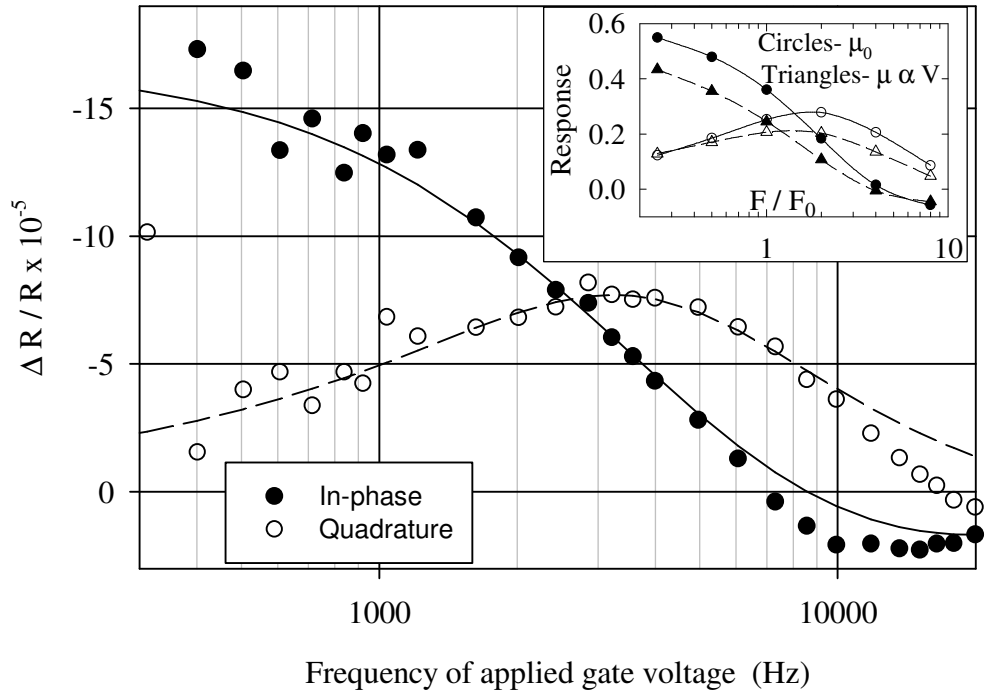

**Figure 2** Change in reflectance for a sample made with alumina dielectric as a function of the gate voltage frequency, $V_{gs}$ = 0 to -100 V. Closed symbols are data in-phase with the gate voltage and open circles are data in quadrature. Satisfactory fits to the solutions of Equation 1 were not possible within the error bars using data at all frequencies; the curves show the fits using the F < 6 kHz data only. Inset shows calculations of response for constant mobility (circles) and charge density dependent mobility ($\mu = \mu_0 V(t)/V_{gs0}$) (triangles.)

### CONCLUSIONS

We find that the frequency dependence of charge flowing into the center of a TIPS-pentacene field-effect transistor can vary within a transistor, and can vary between dielectric types. Individual crystals within one transistor can vary in their transport properties, which are compounded in current-voltage characteristics. Measurements on alumina dielectric show deviation from the calculated charge density response, suggesting that contact impedance affects the transistor at high frequency.

ACKNOWLEDGMENTS

We thank John Anthony for providing TIPS-pentacene and helpful discussions. This work was supported in part by the U.S. National Science Foundation through Grants DMR-0800367 and EPS-0814194 and the Center for Advanced Materials.